\documentclass[bibyear]{aa}

\usepackage{graphicx}

\usepackage{txfonts}

\usepackage{natbib}
\bibpunct{(}{)}{;}{a}{}{,}
\usepackage{booktabs}
\usepackage{multicol}
\usepackage{graphicx}
\usepackage{fancyhdr}

\usepackage{amsmath}	
\usepackage{amssymb}	
\usepackage[inter-unit-product=\cdot]{siunitx}
\usepackage{enumerate}
\usepackage{multirow}

\usepackage{mathtools}

\usepackage{longtable}
\usepackage{tabularx}
\usepackage{xcolor} 
\usepackage{ulem} 
\usepackage{comment}
\usepackage{cuted}

\usepackage{soul}

\usepackage{placeins}
\usepackage{stfloats}
\usepackage{float}

\usepackage{hyperref}
\hypersetup{colorlinks=true,urlcolor=black, citecolor=blue, linkcolor=black}

\makeatletter
\renewcommand*\aa@pageof{, page \thepage{} of \pageref*{LastPage}}
\makeatother

\newcommand{\orcidicon}[1]{\href{https://orcid.org/#1}{\includegraphics[width=11pt]{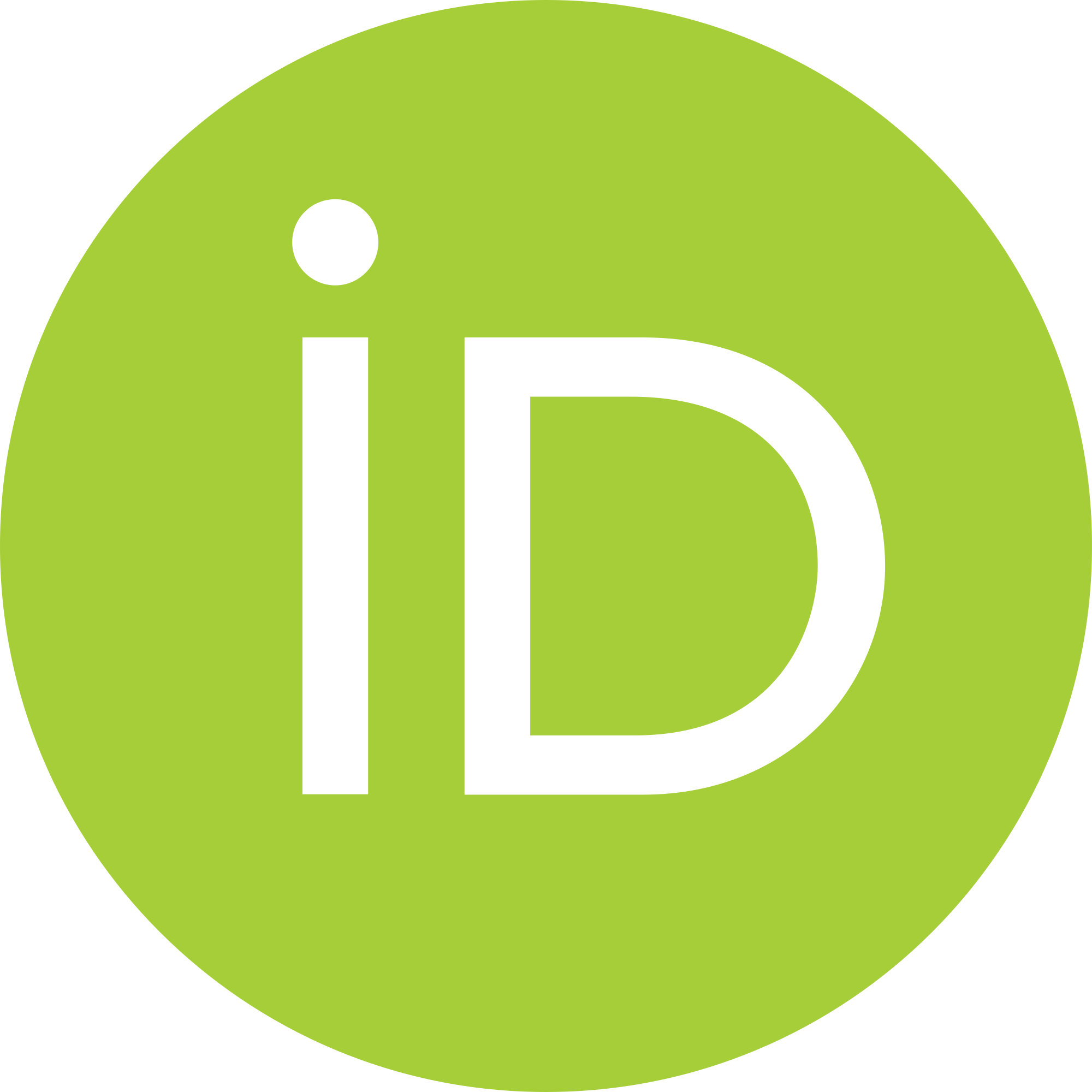}}}
\newcommand{\orcid}[1]{\href{https://orcid.org/#1}{\protect\orcidicon{#1}}}

\definecolor{steelblue}{rgb}{0.274 0.510 0.706}

\begin{document}

   \title{A binary black hole merger rate comparison within the same metallicity--star formation rate  framework}
    \titlerunning{BBH merger rate comparison}

   \author{
    Lumen Boco\inst{1}
    \orcid{0000-0003-3127-922X} \thanks{\href{mailto:lumen.boco@uni-heidelberg.de}{lumen.boco@uni-heidelberg.de}},
    Michela Mapelli\inst{1,2,3,4}
    \orcid{0000-0001-8799-2548},
    Michele Bosi\inst{5,6}
    \orcid{0009-0000-8215-6698},
    Cecilia Sgalletta\inst{1}
    \orcid{0009-0003-7951-4820},
    Amedeo Romagnolo\inst{1,2}
    \orcid{0000-0001-9583-4339}}
    \authorrunning{L. Boco et al.}
    \institute{
    $^{1}$Universit\"at Heidelberg, Zentrum f\"ur Astronomie (ZAH), Institut f\"ur Theoretische Astrophysik, Albert Ueberle Str. 2, 69120, Heidelberg, Germany\\
     $^2$Dipartimento di Fisica e Astronomia Galileo Galilei, Università di Padova, Vicolo dell’Osservatorio 3, I–35122 Padova, Italy\\
    $^3$Universit\"at Heidelberg, Interdiszipli\"ares Zentrum f\"ur Wissenschaftliches Rechnen, D-69120 Heidelberg, Germany\\
    $^4$INFN, Sezione di Padova, Via Marzolo 8, I--35131 Padova, Italy\\
    $^{5}$SISSA, Via Bonomea 365, I–34136 Trieste, Italy\\
    $^6$Department of Physics, University of Trento, Via Sommarive 14, 38123 Povo (TN), Italy\\
    }

   \date{Received XXXX; accepted YYYY}

\abstract{Recent studies have suggested that binary population synthesis models, when coupled with observationally based, metallicity-dependent star formation rate density, overpredict the observed local binary black hole (BBH) merger rate density. The significance of this tension might vary depending on the specific code and parameters adopted. Thus, a more extensive exploration of the parameter space is required. Here, we perform such an extended analysis. We first carry out a comprehensive analysis by using a large set of tabulated BBH merger efficiencies, coming from multiple population synthesis codes across a wide range of physical assumptions, parameters, and initial conditions. We combine these efficiencies with an observationally motivated metallicity distribution and explore 18 variations to encompass observational uncertainties. In this way, we ensure that we compare different population synthesis models under exactly the same assumptions about the metallicity-dependent cosmic star formation rate, which are grounded on observational data. We find that the tension between predicted and observed BBH merger rate density persists: in 17 out of 18 metallicity variations none of the models considered predicts a local BBH merger rate within or below the observed $90\%$ credible interval. Even in the most favorable metallicity variation, only $\lesssim 10\%$ of the models are consistent with the observational constraints. Then, we perform a full calculation, including a self-consistent delay time distribution, for the models yielding the lowest BBH merger rate densities. This self-consistent $t_d$ analysis confirms that even the most conservative models overpredict the observed local BBH merger rate for almost all the metallicity variations. We show that the predicted rates are highly sensitive to the adopted metallicity-dependent star formation history. In particular, the inclusion of the low-metallicity tail contributed by low-mass galaxies and starbursts, together with a treatment of iron rather than oxygen abundance, systematically shift the predicted rates toward higher values. Consequently, even literature models that predict moderate BBH merger rates can instead produce substantially higher rates when coupled to an observationally motivated metallicity distribution. Although not comprehensive of all the literature models, our analysis suggests that models featuring enhanced stellar winds, stronger natal kicks and/or non-standard treatments of mass-transfer and common-envelope physics provide the most promising avenue for alleviating the tension with the observed local BBH merger rate.}


   \maketitle

\defcitealias{Chruslinska2021}{Ch21}
\defcitealias{Chruslinska2025}{Ch25}
\defcitealias{Boco2021}{B21}
\defcitealias{Boco2026}{B26}

\section{Introduction}\label{sec:intro}
The observations of gravitational waves by the LIGO--Virgo--KAGRA (LVK) collaboration over the last ten years have allowed to constrain several properties of the main sources originating them: binary black holes (BBHs). Among the most constrained properties are their masses and merger rates. The Gravitational-Wave Transient Catalog 5.0 \citep[GWTC-5][]{Abbott2026-gwtc5-population} reports a BBH merger rate density at $z=0.2$ of $27.5-49.4\,\textrm{Gpc}^{-3}\,\textrm{yr}^{-1}$ at $90\%$ credible level, while the merger rate at $z=0$ is the same as in GWTC-4: $\sim 15-24\,\textrm{Gpc}^{-3}\,\textrm{yr}^{-1}$. 

Recent works \citep{Sgalletta2025, Boco2026} claimed that the local BBH merger rate density, $\mathcal{R}_0$, predicted by current theoretical models exceeds the observationally inferred value by a factor ${\sim10}$. This is caused by the presence of an extended low-metallicity tail in the metallicity distribution of galaxies (consisting of low-mass galaxies and starbursts), which enhances the BBH merger rate. Furthermore, \citet{Boco2026} (hereafter \citetalias{Boco2026}) showed that considering iron abundance rather than oxygen abundance in deriving the metallicity distribution further exacerbates the discrepancy.

However, these studies rely on few models of stellar and binary evolution, based on a single population synthesis code \citep[\textsc{sevn}, ][]{Iorio2023}. In order to draw more robust conclusions about the theoretically predicted value of $\mathcal{R}_0$, a broader comparison across population synthesis models is necessary, as suggested by \citet{Broekgaarden2026b}. The author presented a comprehensive review of the values of $\mathcal{R}_0$ found by many recent studies, showing that although most models favor merger rates above the observational estimate, several predict values that are consistent with, or even below, the observed rate. 

The value of $\mathcal{R}_0$ is intrinsically related to two quantities: (1) the evolution of the cosmic star formation rate and metallicity distribution, SFRD$(z,Z)$, and (2) the evolution of stars and binaries. These two components are connected via the metallicity that can drastically change the evolution of stellar binaries originating BBHs. Existing studies compute $\mathcal{R}_0$ by using different binary population synthesis codes and different SFRD$(z,Z)$, finding values spanning at least seven orders of magnitude \citep{Mandel2022, Broekgaarden2026a, Broekgaarden2026b}. This large diversity makes it difficult to compare results across studies and to identify the primary sources of the discrepancies. 

In this work, we extend the analysis presented by \citetalias{Boco2026} to multiple studies, standardizing the contribution of the SFRD$(z,Z)$. To this aim, we apply the same observationally motivated SFRD$(z,Z)$ to a broad set of binary population synthesis models. The aim is to marginalize over part of the uncertainties, and to obtain $\mathcal{R}_0$ predictions from different binary population synthesis codes with the same SFRD$(z,Z)$, derived from galaxy observations. As in \citetalias{Boco2026}, we also explore a suite of 18 metallicity variations, corresponding to different parameters of the SFRD$(z,Z)$, bracketing the observational uncertainties in metallicity measurements. This allows us to assess the robustness of our conclusions against uncertainties in the inferred SFRD$(z,Z)$.

We perform two complementary analyses:
\begin{enumerate}
\item A comprehensive analysis, where we combine the SFRD$(z,Z)$ with a large set of BBH merger efficiencies, $\eta(Z)$, compiled by \citet{vanson2025}. This dataset comprises 101 models based on different population synthesis codes and physical assumptions. For this comprehensive analysis, we use the same parametrized delay time distribution, scaling as $t_d^{-1}$, for all the models.
\item A self-consistent $t_d$ analysis, in which we consider only a subset of the available population-synthesis models. Specifically, we select the subset of models yielding the lowest reported BBH merger rate density, i.e. those least in tension with the observations. For these models, we use the self-consistent delay time distributions generated by each population-synthesis model, instead of adopting a generic $t_d^{-1}$ distribution.
\end{enumerate}

The comprehensive analysis confirms that a tension between models and observations is present: for almost all the metallicity variations, 17 out of 18, all the 101 binary evolution models considered are above the $90\%$ credible interval estimated by LVK. Only one metallicity variation features $\lesssim 10\%$ of the binary evolution models within or below such interval. The self-consistent $t_d$ analysis confirms that even the most conservative models overpredict the observed local BBH merger rate for almost all the metallicity variations.

Even literature models that report moderate $\mathcal{R}_0$ values shift towards higher merger rates once a more observationally-calibrated metallicity distribution is assumed. This confirms the importance of considering the low-metallicity tail contributed by low-mass galaxies and starbursts as well as of using iron rather than oxygen when computing the SFRD \citep{Sgalletta2025, Boco2026}. Not including this tail may result in a significant underprediction of the BBH merger rates.

The paper is structured as follows: in Section \ref{sec:methods}, we describe the methods used to compute the BBH merger rate density; in Section \ref{sec:results}, we present and discuss the results; Section \ref{sec:conclusions} summarizes our main findings.

Throughout the paper we assume the Planck2018 cosmology, a flat Universe with $H_0\simeq0.68$ and $\Omega_m\simeq0.31$ \citep{Planck2020}. The reference solar abundances are from \cite{Grevesse1998}, i.e.  $12+\log(\textrm{O}/\textrm{H})_\odot=8.83$, $12+\log(\textrm{Fe}/\textrm{H})_\odot=7.5$, $\log(\textrm{O}/\textrm{Fe})_\odot=1.33$, and $Z_\odot=0.017$.

\section{Methods}\label{sec:methods}
We compute the BBH merger rate density at the cosmic time $t$ as \citep{Boco2019, Boco2021, Neijssel2019,  Santoliquido2022}:
\begin{equation}
\mathcal{R}(t)=\int\,\textrm{d}\log Z\,\eta(Z)\int\textrm{d}t_d\,\frac{\textrm{d}p}{\textrm{d}t_d}\,\textrm{SFRD}(t-t_d,Z)
\label{eq:merger_rate}
\end{equation}
where $\eta(Z)$ is the BBH merger efficiency as a function of metallicity $Z$ and $\textrm{d}p/\textrm{d}t_d$ is the distribution of the delay time between the formation of the binary and the merger of the remnant black holes. These two terms come from stellar and binary evolution and are usually computed through binary population synthesis simulations. The SFRD is the metallicity-dependent star formation rate density of the Universe, calculated at the binary formation time $t-t_d$, and it is usually either computed from cosmological simulations \citep{Mapelli2017, Oshaughnessy2017, Lamberts2018, Mapelli2018, Artale2019, Levina2026} or inferred from observations \citep{Belczynski2016, Lamberts2016, Cao2018, Elbert2018, Li2018, Boco2019, Chruslinska2019, Neijssel2019, Santoliquido2020, Boco2021, Santoliquido2022, Broekgaarden2022, Romagnolo2023,Romagnolo2025,Sgalletta2025}. In the next subsections we describe how we derive these two terms. 

\subsection{Stellar and binary evolution}
\subsubsection{Comprehensive analysis}
In the comprehensive analysis, we compute the local BBH merger rate density $\mathcal{R}_0$ for a large set of binary evolution models available in literature. We take the BBH merger efficiencies $\eta(Z)$ from the compilation of \citet{vanson2025}\footnote{We derive the BBH merger efficiencies using the online material at \href{https://zenodo.org/records/14508864}{https://zenodo.org/records/14508864}.}. This is a collection of 101 models coming from different widely used binary population synthesis codes -- \textsc{sevn} \citep{Spera2017, Spera2019, Mapelli2020, Iorio2023}, \textsc{mobse} \citep{Mapelli2017,Giacobbo2018,Santoliquido2021}, \textsc{compas} \citep{Stevenson2017:compas,Barrett2018:compas,VignaGomez2018:compas,Broekgaarden2019:compas,Neijssel2019:compas,Broekgaarden2022}, and \textsc{startrack} \citep{Belczynski2002,Belczynski2008, Chruslinska2018, Klencki2018, Belczynski2020} -- with different physical assumptions and parameters. Table \ref{table} provides a summary of the models considered and their respective references for more detail.

\begin{table}
\centering
\footnotesize 
\begin{tabular}{p{0.4\columnwidth}  p{0.17\columnwidth}  p{0.32\columnwidth}}
\hline
Reference & Code & Model names \\
\hline
\citet{Iorio2023} & SEVN & F, F5M, QCBSE, QCBB, NT, NTC, RBSE, SND, K$\sigma$150, K$\sigma$265, LX, LK, LC, OPT, QHE, F19  \\

\citet{Broekgaarden2022} & COMPAS &
A, B, C, D, E, F, G, H, I, J, K, L, M, N, O, P, Q, R, S, T  \\

\citet{Neijssel2019} &
COMPAS &
Preferred model\\

\citet{Giacobbo2018} &
MOBSE &
$\alpha1$, $\alpha3$, $\alpha5$, CC15$\alpha1$, CC15$\alpha3$, CC15$\alpha5$\\

\citet{Santoliquido2021} &
MOBSE &
$\alpha0.5$, $\alpha1$, $\alpha2$, $\alpha3$, $\alpha5$, $\alpha7$, $\alpha10$\\

\citet{Klencki2018} &
STARTRACK &
M10, I1, I2\\

\hline
\end{tabular}
\vspace{+1mm}
\caption{Summary of the binary evolution models considered in this work. First column: reference to the original work. Second column: population synthesis code name. Third column: Name of the model, as used in the original work. Note that, for \citet{Iorio2023}, every model actually corresponds to four models computed at different common envelope efficiencies: $\alpha=0.5,\,1,\,3,\,5$.}
\label{table}
\end{table}

These models explore the evolution of binaries under a variety of initial conditions (initial masses, binary fraction, orbital period, and eccentricity) and physical assumptions (stellar evolution tracks, stellar winds, mass transfer and common envelope physics, supernova physics and natal kicks). Despite the diversity of codes and models considered, we caution that they do not represent a complete sample of different codes, assumptions and parameters present in the literature, but only those models whose BBH merger efficiency was included in \citet{vanson2025}. Moreover, some population synthesis codes and sets of physical parameters may be more extensively represented than others. Thus, the $\mathcal{R}_0$ distributions and the percentages reported in this work refer specifically to the set of models included in this study.

For this comprehensive analysis, we adopt a common delay time distribution for all models, scaling as $\propto t_d^{\alpha_D}$, with minimum delay time $t_{d,\textrm{min}}=3$ Myr, and with $\alpha_D=-1$. Despite many studies report delay time distributions consistent with $t_d^{-1}$, this may result in some inconsistencies of our calculated merger rate densities with the original simulations, as the merger efficiency and the delay time distribution are mutually dependent. In addition, the delay time distribution may depend on metallicity \citep{Fishbach2023, Boesky2024a}. The choice to adopt a common delay time distribution in the comprehensive analysis is done for computational simplicity: obtaining a self-consistent delay time distribution for all 101 models would require access to, and analysis of, the full simulation outputs of each population-synthesis model. To assess the sensitivity of our results to this assumption, we also repeat the calculation for $\alpha_D=-0.5$ and $-1.5$, in addition to our fiducial choice $\alpha_D=-1$.

\subsubsection{Self-consistent $t_d$ analysis}\label{sec:self-consistent}
In the self-consistent $t_d$ analysis, we re-calculate the merger rate density of a subset of the models already included in the comprehensive analysis, using  their original delay time distributions instead of the simplified power-law assumption. The purpose of our self-consistent $t_d$ analysis is thus to show the impact of a more complex delay time distribution on the final merger rate density estimation.

We select the subset of models yielding the
lowest reported BBH merger rate density, i.e. those least
in tension with the observations. We consider nine models: the QHE, QCBSE, and K$\sigma 265$ models by \citet{Iorio2023} with common envelope efficiency $\alpha_{\textrm{CE}}=5$; the $\alpha0.5$, $\alpha7$ and $\alpha10$ models by \citet{Santoliquido2021}; and the models G, J, and T by \citet{Broekgaarden2022}. The three selected \textsc{sevn} models \citep{Iorio2023} adopt a quasi-homogeneous evolution scenario (QHE), the BSE-like mass-transfer stability criteria (QCBSE), and a scenario in which natal kicks are directly drawn from a Maxwellian distribution with $\sigma_\textrm{kick}=265\ \textrm{km s}^{-1}$ and no fallback modulation (K$\sigma 265$). The $\alpha0.5$, $\alpha7$, and $\alpha10$ models of \textsc{mobse} \citep{Santoliquido2021} differ only by the value of the common-envelope efficiency parameter: $\alpha_\textrm{CE}=0.5$, $\alpha_\textrm{CE}=7$ and $\alpha_\textrm{CE}=10$, respectively. These three $\alpha_\textrm{CE}$ values yield particularly low BBH merger efficiencies. The G and J \textsc{compas} models \citep{Broekgaarden2022} adopt $\alpha_\textrm{CE}=0.1$, $\alpha_\textrm{CE}=10$, respectively; while model T (which yields the lowest BBH merger efficiency) features Wolf-Rayet\footnote{SSE-based models (e.g., \textsc{COMPAS, MOBSE, StarTrack}) originally referred only to "naked helium stars" \citep{Hurley_2000} and do not have any evolutionary prescription for hydrogen-rich Wolf-Rayet objects (also known as WNh). Most SSE-based models in the recent literature now refer to Wolf-Rayet stars as a synonym for naked helium stars. Here we refer to Wolf-Rayet stars to keep the notation consistent with \citep{Broekgaarden2022}.} winds multiplied by a factor $5$ compared to the default model.

For these nine models, we directly take the outputs of the corresponding population synthesis simulations to recompute the BBH merger efficiency and the delay time distribution\footnote{We take \textsc{sevn} outputs from \href{https://zenodo.org/records/7794546}{https://zenodo.org/records/7794546}, and \textsc{compas} outputs from \href{https://zenodo.org/records/5651073}{https://zenodo.org/records/5651073}. The \textsc{mobse} outputs were shared by the authors of \cite{Santoliquido2021}.}.

\subsection{Metallicity-dependent cosmic star formation rate density (SFRD)}
As for the metallicity-dependent SFRD$(z,Z)$, we compute it in two different ways: (i) the fiducial configuration, and (ii) the symmetric configuration. This is done to show the effect of using a simple symmetric metallicity distribution, with respect to an observationally-derived one. 

\subsubsection{The fiducial configuration}\label{sec:fiducial}
In the fiducial configuration, we compute the SFRD$(z,Z)$ as in equation (2) of \citetalias{Boco2026} \citep[see also][]{Boco2021}, using galaxy observed statistics and empirical scaling relations: 
\begin{equation}
\begin{split}
\textrm{SFRD}(z,Z)\equiv\frac{\textrm{d}\dot M}{\textrm{d}V\,\textrm{d}\log Z}(z,Z)=\int_{\log M_{\star, \textrm{min}}}\textrm{d}\log M_\star\,\frac{\textrm{d}N}{\textrm{d}V\,\textrm{d}\log M_\star}\\\int\textrm{d}\log\psi\,\psi\, \frac{\textrm{d}p}{\textrm{d}\log\psi}(\log\psi|z,M_\star)\,\frac{\textrm{d}p}{\textrm{d}\log Z}(\log Z|M_\star,\psi)
\end{split}
,
\end{equation}
where $\textrm{d}N/\textrm{d}V\,\textrm{d}\log M_\star$ is the star-forming galaxy stellar mass function, $M_{\star, \textrm{min}}$ is the minimum stellar mass accounted for in the integration, $\psi$ is the star formation rate (SFR), $\textrm{d}p/\textrm{d}\log\psi$ is the SFR distribution at a given stellar mass and redshift, and $\textrm{d}p/\textrm{d}\log Z$ is the metallicity distribution as a function of the galaxy stellar mass and SFR, which is gaussian in $\log Z$ centered around an average value given by the fundamental metallicity relation (FMR), with scatter $\sigma_\textrm{FMR}$. We adopt the galaxy stellar mass function by \citet{Weaver2023} with the main sequence of \citet{Popesso2023}, and the FMR by \citet{Chruslinska2021}, with three free parameters, $Z_{\textrm{O/H},0}$, $a_{\textrm{MZR}}$, and $\nabla_{\textrm{FMR},0}$, to account for uncertainties in metallicity measurements. 

Specifically, $Z_{\textrm{O/H},0}$ is the normalization of the FMR, and $a_{\textrm{MZR}}$ is the slope of the mass-metallicity relation at redshift $z=0$. If $a_{\textrm{MZR}}\sim{0}$ it means that the metallicity of a galaxy is not affected by its stellar mass, whereas a large value of $a_{\textrm{MZR}}$ indicates a strong relation between mass and metallicity (i.e., low-mass galaxies are more metal poor than high-mass galaxies). Finally $\nabla_{\textrm{FMR},0}$ is the slope of the anticorrelation between metallicity and star formation rate.  
We refer to \citetalias{Boco2026} for additional information on these parameters. As in \citetalias{Boco2026}, we perform 18 \textit{metallicity variations} by varying these parameters in the ranges: $Z_{\textrm{O/H},0}=8.8,\,9.0,\,9.2$, $a_{\textrm{MZR}}=0.15,\,0.3,\,0.6$, and $\nabla_{\textrm{FMR},0}=0.2,\,0.3$, which encompass many observational metallicity determinations in the literature \citep{Lara-Lopez2010, Mannucci2010, Mannucci2011, Andrews2013, Zahid2014, Hunt2016, Cresci2019, Curti2020, Sanders2021, Curti2023, Nakajima2023}.

We also account for the $\textrm{O/Fe}$ correction by \citet{Chruslinska2025} to pass from oxygen abundance to iron abundance. This is done because binary evolution is more sensitive to iron rather than oxygen, as iron group elements set the strengths of line-driven winds in massive stars, at least for $Z\gtrsim0.1~Z_\odot$ \citep{Romagnolo_2026_AtlasI}, and therefore can widely affect stellar expansion \citep[e.g.][]{Agrawal2020, Romagnolo2023,vanson2025}, final BH masses \citep{Kruckow2024,Romagnolo2024,Merritt2026,Romagnolo2026b, Boco2025}, and consequently binary evolution.

Since \citetalias{Boco2026} wanted to minimize the BBH merger rates, they did several conservative assumptions on the SFRD calculation. Here, in order to get more realistic results, we relax some of these assumptions. The main differences are three:

\begin{itemize}
\item We integrate down to galaxy stellar mass $M_{\star,\textrm{min}} = 10^7\,M_\odot$, rather than $M_{\star,\textrm{min}} = 10^8\,M_\odot$;
\item We use the estimate of \citet{Chruslinska2025} to set the fraction of starburst galaxies $f_b$, rather than setting $f_{\textrm{SB}}=0$, as done in \citetalias{Boco2026}. The starburst sequence is shifted by $1$ dex above the main sequence for $M_\star>10^9\,M_\odot$, and it decreases linearly from $1$ dex to $0$ dex for $10^6\,M_\odot<M_\star<10^9\,M_\odot$.
\item We set a scatter around the fundamental metallicity relation $\sigma_{\textrm{FMR}}=0.1$ rather than $\sigma_{\textrm{FMR}}=0.05$ to keep into account further scatter in the $\textrm{O}/\textrm{Fe}$ relation and the internal scatter within the same galaxy.
\end{itemize}
In Section \ref{sec:starbursts}, we will also show a comparison between the fiducial and a conservative configuration where the contribution of the starburst galaxies is ignored ($f_\textrm{SB}=0$) to show the importance of including starburst galaxies in the calculation of BBH merger rates. We also present results for the \citetalias{Boco2026} configuration, with $f_\textrm{SB}=0$, $M_{\star,\textrm{min}} = 10^8\,M_\odot$, and $\sigma_{\textrm{FMR}}=0.05$, which has to be intended as the most conservative configuration, i.e. the one yielding the lowest BBH merger rate density values according to current BBH models \citepalias[see the extensive discussion in Section 2 of][]{Boco2026}.

\begin{figure*}
    \centering
    \includegraphics[width=1.0\linewidth]{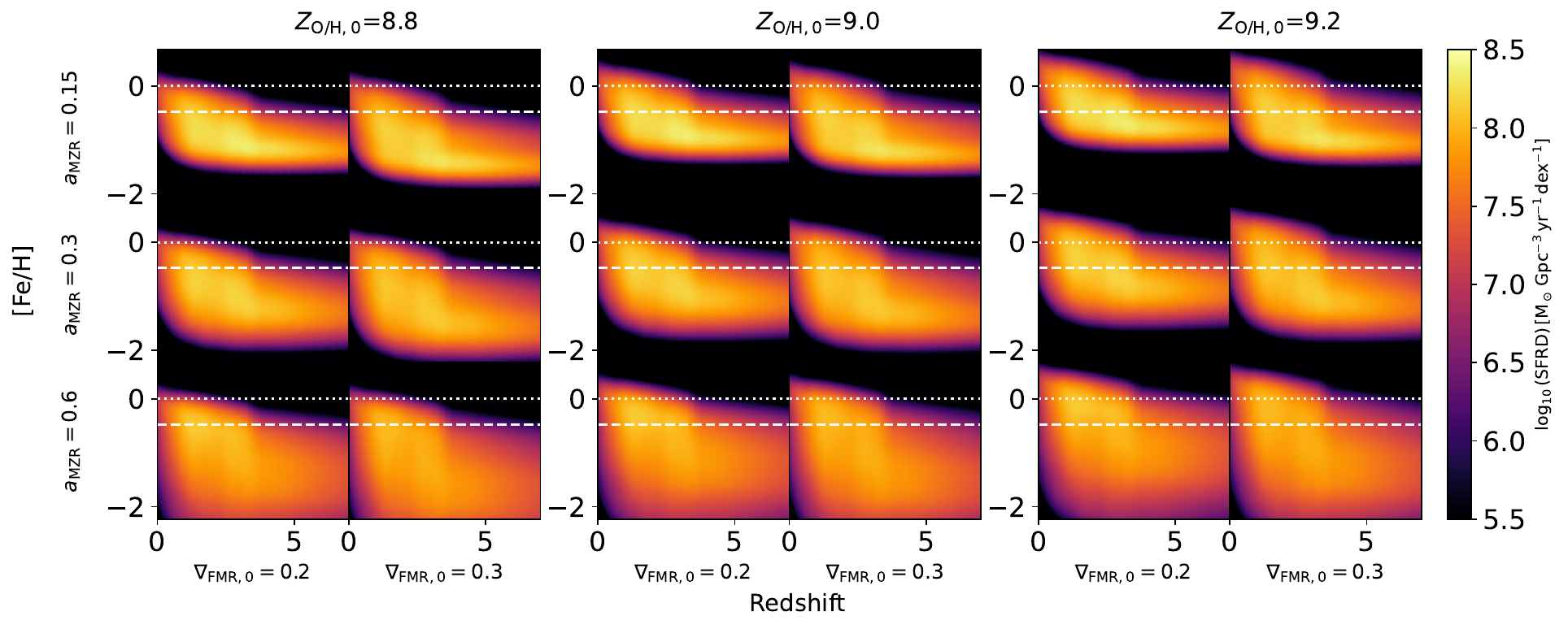}
    \caption{Cosmic SFRD per metallicity bin, for all the metallicity variations considered within the fiducial configuration, as a function of $z$ and [Fe/H]. Different panels are for different values of $Z_{\textrm{O/H},0}$, $a_{\textrm{MZR}}$ and $\nabla_{\textrm{FMR},0}$. The white dashed and dotted horizontal lines represent $Z_\odot/3$ and $Z_\odot$.}
    \label{fig:SFRD}
\end{figure*}

Figure \ref{fig:SFRD} shows the SFRD$(z,Z)$ in the fiducial configuration for the different metallicity variations, corresponding to the different values of the parameters $Z_{\textrm{O/H},0}$, $a_{\textrm{MZR}}$, and $\nabla_{\textrm{FMR},0}$. The metallicity distribution is not symmetric in $\log Z$, and the low-metallicity tail increases for larger $a_{\textrm{MZR}}$ and $\nabla_{\textrm{FMR},0}$, and for lower $Z_{\textrm{O/H},0}$, as shown in \citetalias{Boco2026}. Since the BBH merger efficiency strongly depends on metallicity, the prominence and extension of this low-metallicity tail is fundamental to determine the BBH merger rate density.

\subsubsection{The symmetric configuration}
Since we want to show the impact of the choice of the metallicity distribution on the BBH merger rates, we compare our fiducial configuration with a popular choice in this type of studies: a model with a mean redshift-dependent metallicity, with a symmetric scatter in $\log Z$ around it, $\sigma_Z$. The average metallicity is taken from \citet{Madau2017}, who derive a mean mass-weighted metallicity at each redshift from observations of oxygen abundance $\log \langle Z/Z_\odot\rangle=0.153-0.074\,z^{1.34}$.

The scatter in the symmetric configuration is then expressed using a lognormal distribution
\begin{equation}
    p(Z \vert z) = \frac{1}{\sqrt{2 \pi {\sigma}_{\rm Z}^2}} \exp \left[ -\frac{[\log Z/Z_\odot - \mu(z)]^{2}}{2 {\sigma}_Z^{2}}\right]
\end{equation}
where $\mu(z)=\langle \log{Z/Z_\odot}\rangle{}=\log{\langle{}Z/Z_\odot\rangle{}} - {\ln{(10)}}\,{}\sigma_Z^2/2$. As the values of the scatter, we choose $\sigma_Z=0.1,\,0.2,\,0.3,\,0.6$ to show the severe impact of the choice of the scatter on the BBH merger rates. Note that $\sigma_Z$ is different from $\sigma_\textrm{FMR}$. $\sigma_Z$ of the symmetric configuration accounts for the overall metallicity dispersion at a fixed redshift, while $\sigma_\textrm{FMR}$ renders the scatter around the FMR, i.e. the metallicity dispersion for galaxies with the same stellar mass and star formation rate. The overall scatter at a given redshift in the fiducial configuration is way larger than $\sigma_\textrm{FMR}$, and it is driven by the metallicities of individual galaxies with different $M_\star$ and star formation rate. 

In the symmetric configuration, we adopt the same cosmic star formation rate density as in the fiducial case, so as to isolate the effect of changing the metallicity distribution while keeping the overall star formation history fixed. This decoupling between star formation rate and metallicity distribution is a common assumption in literature. However, this approach may be affected by some biases. Since we are interested in the metallicity of newly formed binaries, the contributions from individual galaxies should be weighted by their star formation rate. This weighting is naturally incorporated in our fiducial configuration, which therefore provides a more consistent description of the metallicity distribution of newly formed binaries.

Table \ref{table1} reports the two configurations considered with their associated parameters.

\begin{table}
\centering
\footnotesize 
\begin{tabular}{p{0.19\columnwidth}  p{0.38\columnwidth}  p{0.32\columnwidth}}
\hline
Configuration & Description & Parameters \\
\hline
Fiducial\newline configuration & Fundamental metallicity relation + iron correction & $Z_{\textrm{O/H},0}=8.8,\,9.0,\,9.2$,\newline $a_{\textrm{MZR}}=0.15,\,0.3,\,0.6$,\newline $\nabla_{\textrm{FMR},0}=0.2,\,0.3$  \\[8mm]

Symmetric configuration & Average metallicity + symmetric scatter in $\log Z$ &
$\sigma_Z=0.1,\,0.2,\,0.3,\,0.6$  \\

\hline
\end{tabular}
\vspace{+1mm}
\caption{The table report the two configurations considered, a brief description, and the parameters associated to them.}
\label{table1}
\end{table}

\section{Results}\label{sec:results}
\subsection{Adopting an observationally-calibrated SFRD induces a  tension between models and data} \label{sec:results_1}
\begin{figure*}
    \centering
    \includegraphics[width=1.0\linewidth]{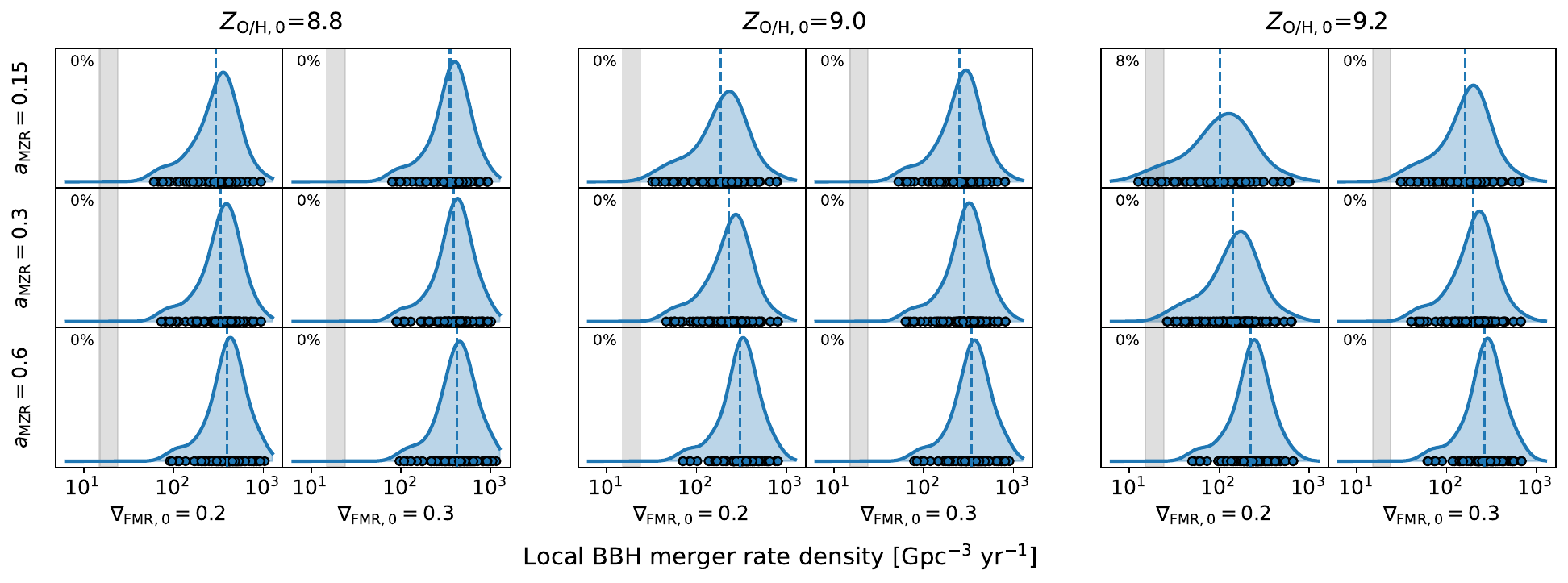}
    \caption{ BBH merger rate density at redshift zero $\mathcal{R}_0$ predicted by different binary evolution models, for all the metallicity variations considered. The blue circles represent different models, the blue line and shaded area a distribution of $\mathcal{R}_0$ given by different binary evolution models, the blue vertical dashed line the mean $\mathcal{R}_0$ from all the models. The gray vertical area is the $90\%$ credible interval for $\mathcal{R}_0$ estimated by LVK \citep{Abbott2026-gwtc5-population}. The percentage reported in the top left of each panel shows the fraction of models falling within or below the LVK $90\%$ credible interval.} 
    \label{fig:main}
\end{figure*}

Figure \ref{fig:main} shows the results of the comprehensive analysis over the 101 considered binary evolution models\footnote{The kernel density estimation is done only for visual reference, to better understand where most of the models considered here cluster.} The figure shows the local BBH merger rate $\mathcal{R}_0$ predicted by the different models, for the 18 metallicity variations considered in the fiducial configuration. In 17 out of 18 metallicity variations, all models fall above the $90\%$ credible interval by LVK. In the last most favorable metallicity variation, a small percentage of models ($\lesssim 10\%$) fall inside or below the $90\%$ credible interval.

The Figure shows that an observationally-derived SFRD implies a tension between models' predictions and data. This extends the results of \citetalias{Boco2026} to a much wider range of binary evolution models and population synthesis codes. We stress that many of these models were associated to lower $\mathcal{R}_0$ in the original works and in the compilation by \citet{Broekgaarden2026b}.

The reason for these differences is that the different studies adopt different SFRD$(z,Z)$ to compute $\mathcal{R}_0$. In this work, instead, we standardize the star formation and metallicity relation for all the different models, and, most importantly, we anchor metallicity to observational galaxy scaling relations. We account for the low-metallicity tail originated by low-mass galaxies and we consider iron abundance rather than oxygen abundance. These differences in the treatment of the SFRD$(z,Z)$ increase the estimated $\mathcal{R}_0$ compared to the original works, so exacerbating the tension between model predictions and observations. 

\begin{figure}
    \centering
    \includegraphics[width=1.0\linewidth]{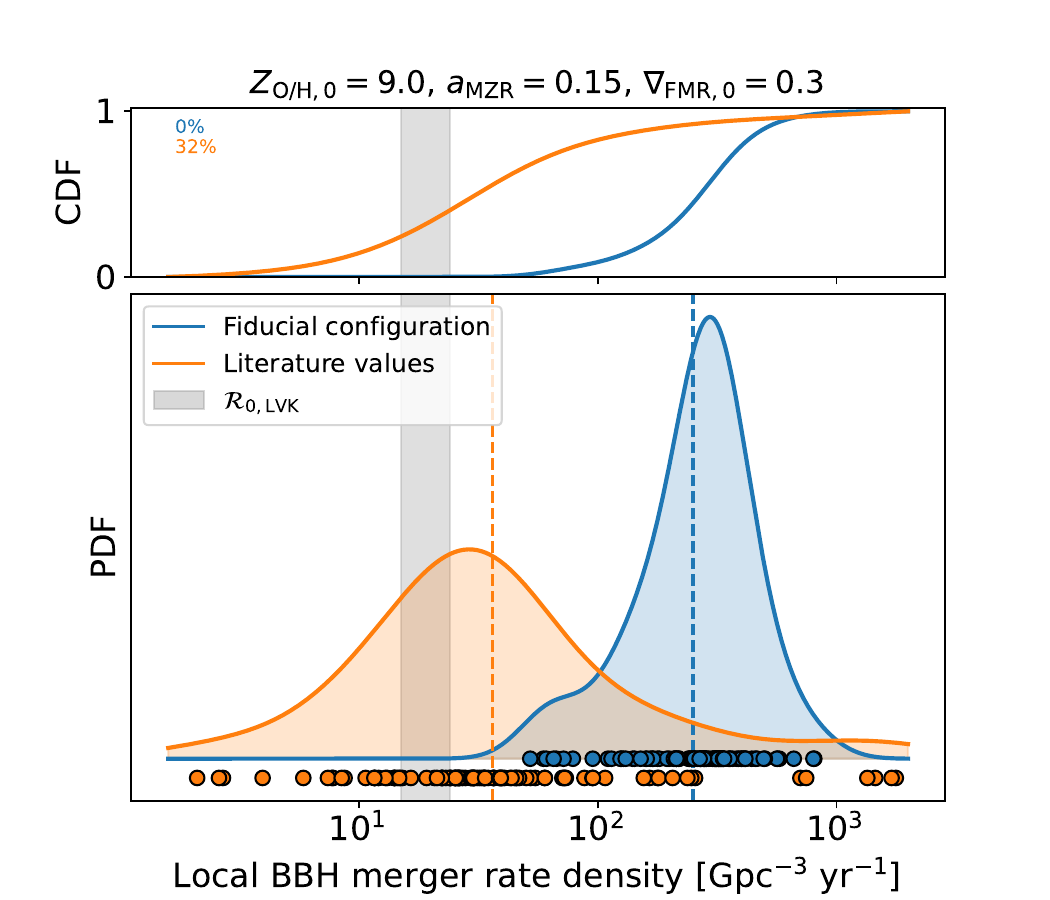}
    \caption{Comparison between $\mathcal{R}_0$ values derived in our fiducial SFRD configuration, for $Z_{\textrm{O}/\textrm{H},0}=9.0$, $a_{\textrm{MZR}}=0.15$, and $\nabla_{\textrm{FMR},0}=0.3$ (blue points and shaded area) and $\mathcal{R}_0$ taken from the original works at face value (orange points and shaded area).} 
    \label{fig:literature}
\end{figure}

Figure \ref{fig:literature} shows this effect in detail. We compare our derived $\mathcal{R}_0$ values, for the metallicity variation $Z_{\textrm{O}/\textrm{H},0}=9.0$, $a_{\textrm{MZR}}=0.15$, $\nabla_{\textrm{FMR},0}=0.2$, and the $\mathcal{R}_0$ values reported in the original works, taken at face value. The original works use different functional forms of the SFRD$(z,Z)$. Specifically, \citet{Klencki2018, Neijssel2019, Santoliquido2021} and \citet{Iorio2023} use a symmetric distribution in $\log Z$ (equivalent to the symmetric configuration of this work), with $\sigma_Z=0.5,\,0.39,\,0.2,\,0.2$, respectively. \citet{Giacobbo2018} assume all stars in a given redshift bin to have the same metallicity, which is equivalent to $\sigma_Z=0$. \citet{Broekgaarden2022}, instead, calculate $\mathcal{R}_0$ for different SFRD, which are computed by convolving different galaxy stellar mass functions with various mass-metallicity relations. Here, we adopt their fiducial model, which is the same as \citet{Neijssel2019}.

The use of our observationally-calibrated SFRD (which is a significant improvement compared to old approximations based on the average cosmic star formation rate) increases the tension between models and observations. Compiling $\mathcal{R}_0$ values directly from the literature yields a broad $\mathcal{R}_0$ distribution, with $\sim 32\%$ of the models within or below the LVK $90\%$ credible interval. Instead, when a SFRD with extended low-metallicity tail is used, all the models predict $\mathcal{R}_0$ above the LVK $90\%$ credible interval.

\begin{figure}
    \centering
    \includegraphics[width=1.0\linewidth]{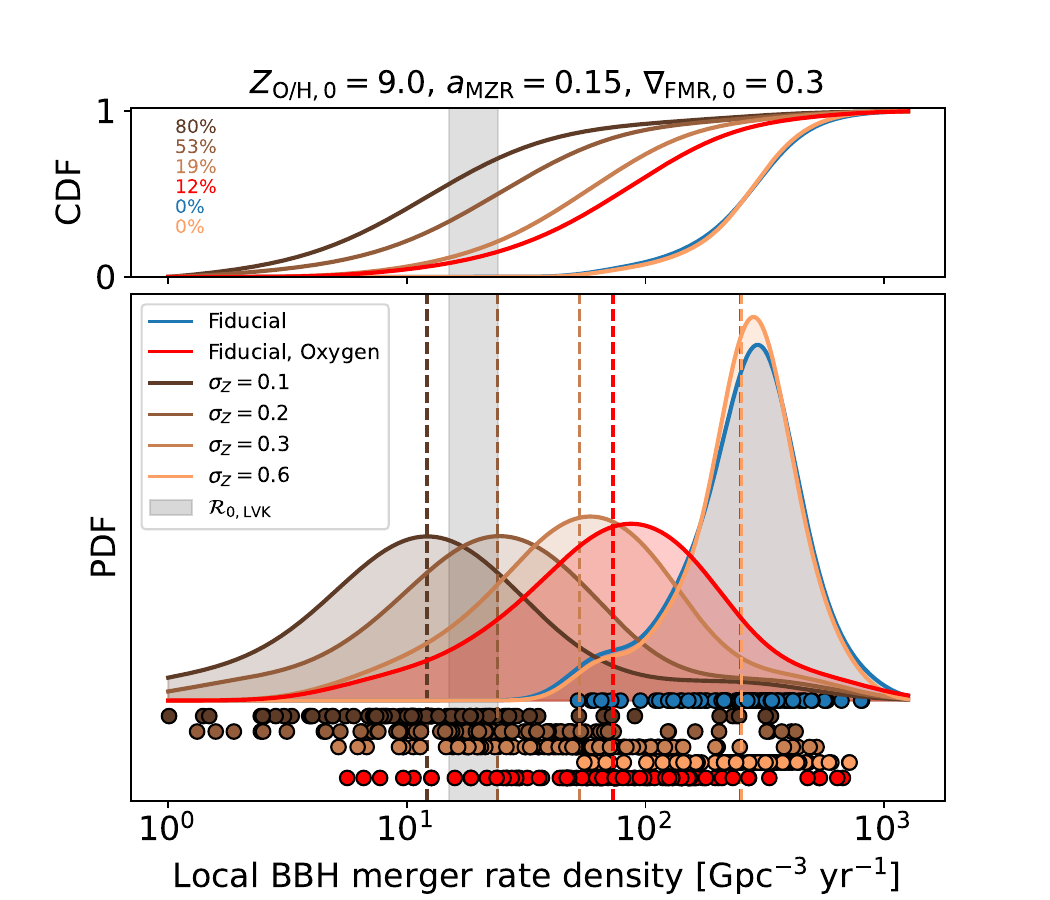}
    \caption{Circles, shaded areas and vertical dashed lines represent $\mathcal{R}_0$ values, distribution and mean, as in Figure \ref{fig:main}. Different colors show different metallicity distributions. The blue color represents our fiducial SFRD configuration, in the metallicity variation $Z_{\textrm{O}/\textrm{H},0}=9.0$, $a_{\textrm{MZR}}=0.15$, and $\nabla_{\textrm{FMR},0}=0.3$. Red represents our fiducial configuration but with no iron correction, thus showing the effect of considering oxygen abundance. The brown scale represents a symmetric metallicity distribution with scatter $\sigma_Z=0.1,\,0.2,\,0.3,\,0.6$. Only models adopting symmetric metallicity distributions have large overlap with the LVK 90\% credible interval.}
    \label{fig:sigma}
\end{figure}

To demonstrate the impact of the low-metallicity tail on the predicted BBH merger rates, Figure \ref{fig:sigma} compares the $\mathcal{R}_0$ distribution from our fiducial configuration, for $Z_{\textrm{O}/\textrm{H},0}=9.0$, $a_{\textrm{MZR}}=0.15$, and $\nabla_{\textrm{FMR},0}=0.3$, with the symmetric configuration with different $\sigma_Z$ values. We also show the case where no iron correction is adopted and oxygen abundance is used as a proxy for metallicity. The choice of $\sigma_Z$ has a strong impact on the predicted $\mathcal{R}_0$ distribution. For $\sigma_Z=0.1,\,0.2,\,0.3,\,0.6$, the percentage of models with $\mathcal{R}_0$ within or below the LVK $90\%$ credible interval is $80\%,\,53\%,\,19\%,\,0\%$, respectively. For our fiducial configuration, instead, all the models overestimate the local BBH merger rates. This clearly shows the difference between a metallicity distribution that is symmetric in $\log Z$ around a mean value and a distribution featuring an extended low-metallicity tail. Only symmetric distributions with $\sigma_Z\gtrsim 0.6$ can mimic the presence of such a tail.

The iron correction has an additional substantial impact. Including the correction shifts the predicted merger-rate distribution toward higher values, increasing its mean by a factor of $\sim 3$, in agreement with \citet{Chruslinska2025}, while also making the distribution narrower. When no iron correction is introduced, $12\%$ of the models are within or below the 90\% credible interval inferred by the LVK collaboration.

\subsection{Additional variations of the star formation rate density}\label{sec:starbursts}
In the previous Section, we have explored several variations of the SFRD. We considered 18 variations of the FMR parameters, spanning a wide range of normalizations, mass-metallicity slopes, and strengths of the anti-correlation between star formation rate and metallicity. We have also shown the effect of using oxygen rather than iron abundance. In this Section we explore the effect of variations in the SFRD beyond the FMR parameters. In particular, we investigate more conservative assumptions for the starburst fraction, the minimum galaxy stellar mass, and the scatter around the FMR. All the variations were chosen in the direction of reducing the predicted $\mathcal{R}_0$, thus providing conservative tests of the tension between the predicted and observed BBH merger rates.

\begin{figure*}
    \centering
    \includegraphics[width=1.0\linewidth]{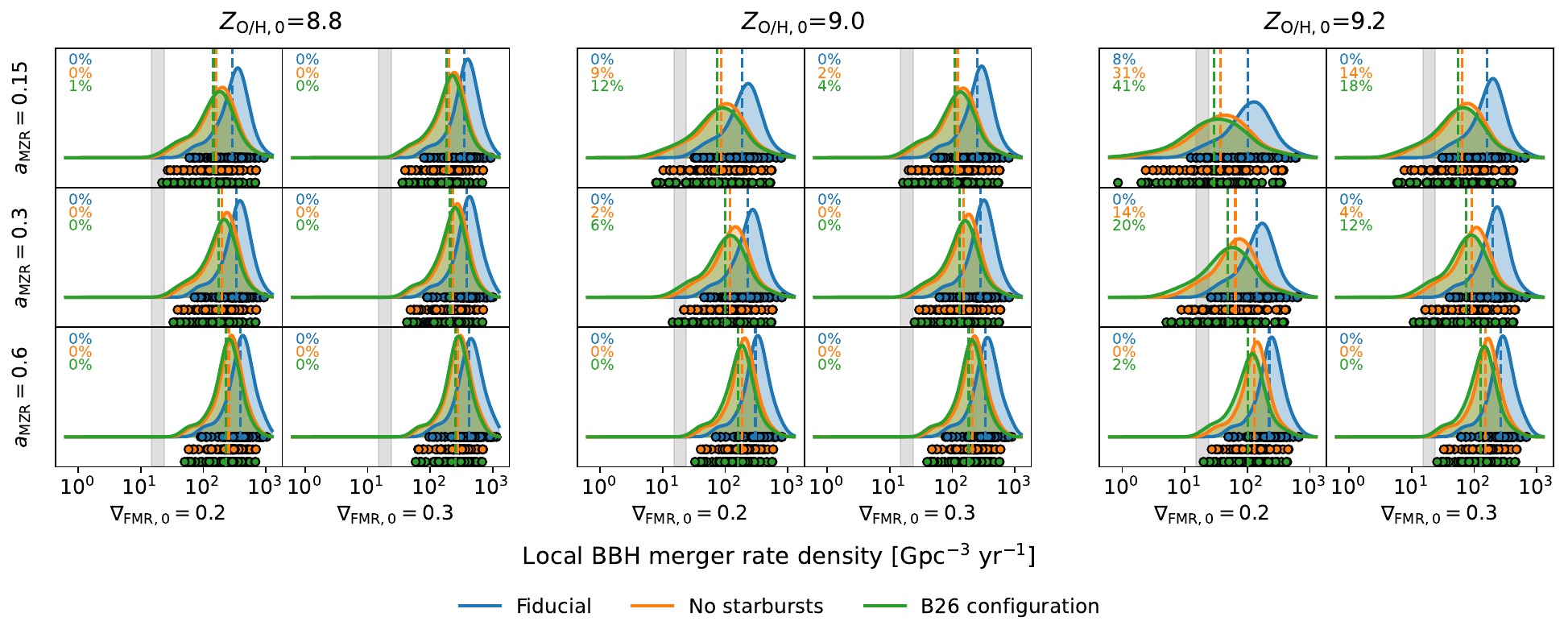}
    \caption{Circles, shaded areas and vertical dashed lines represent $\mathcal{R}_0$ values, distributions and means, as in Figure \ref{fig:main}. Different colors show different configurations. Blue represents our fiducial configuration, orange represents the fiducial configuration but with starburst fraction set to $f_{\textrm{SB}}=0$, green is the B26 configuration, with no starbursts, minimum galaxy mass $M_{\star,\textrm{min}}=10^8\,M_\odot$, and $\sigma_{\textrm{FMR}}=0.05$.}
    \label{fig:main_B26}
\end{figure*}

Figure \ref{fig:main_B26} compares our fiducial configuration with a configuration in which starburst galaxies are excluded, $f_\textrm{SB}=0$, and with the most conservative configuration of \citetalias{Boco2026}, which additionally increases the minimum galaxy mass to $M_{\star,\textrm{min}}=10^8\,M_\odot$ and reduces the scatter around the FMR to $\sigma_{\textrm{FMR}}=0.05$. Removing starburst galaxies has an appreciable effect, making the rate distribution shallower and shifted to lower $\mathcal{R}_0$. The magnitude of this effect depends strongly on the metallicity variation. For variations with lower average metallicities the effect is small, while for variations with higher average metallicities the contribution of starbursts becomes important, since they are the main contributors to the low-metallicity tail. Even in the no starburst case, however, a tension with the observed BBH merger rate remains, although it is alleviated. For the majority of the metallicity variations (11 out of 18) all binary evolution models still predict rates above the observations. For 6 variations, a small percentage of models ($\sim 2\%-14\%$) falls within or below the LVK 90\% credible interval, increasing to $31\%$ for the most favorable metallicity variation.

This result further highlights the importance of including starburst galaxies when constructing the metallicity-dependent SFRD: omitting them leads to substantially lower BBH merger rates than those obtained with a more complete SFRD model. At the same time, the starburst population remains subject to significant observational uncertainties \citep{Rodighiero2011, Sargent2012, Schreiber2015, Caputi2017, Bisigello2018, Boco2021, Chruslinska2025, Rinaldi2025}, making it one of the major uncertainties in the modeling of the galaxy component of BBH merger rate calculations. The no starburst case presented here should therefore not be regarded as a realistic description of the cosmic star formation history,  but rather as a limiting case providing a lower bound for the variations of the SFRD.

By increasing the minimum galaxy mass, from $M_{\star,\textrm{min}}=10^7\,M_\odot$ to $M_{\star,\textrm{min}}=10^8\,M_\odot$, and the scatter around the FMR, from $\sigma_{\textrm{FMR}}=0.1$ to $\sigma_{\textrm{FMR}}=0.05$, we recover the most conservative results presented in \citetalias{Boco2026}. This further alleviates the tension, but the effect of $M_\star{}$ and $\sigma_{\textrm{FMR}}$ is minor when compared to starbursts.

Other possible sources of uncertainty of the SFRD concern the evolution of the FMR and its scatter with redshift. Here, we adopted a redshift independent FMR, since recent studies \citep[e.g.][]{Nakajima2023} showed an invariance of the FMR up to $z\sim3-7$, though with a possible increase in scatter, depending on the adopted parametrization. An FMR declining with redshift or a scatter increasing with redshift would enhance the contribution of low-metallicity star formation and therefore increase the predicted BBH merger rate density. Thus we do not explore these variations in this work. Finally, variations of the galaxy star-forming main sequence have already been studied in \citet{Sgalletta2025}.

\subsection{The effect of the delay time distribution}\label{sec:delay}
\begin{figure*}
    \centering
    \includegraphics[width=1.\linewidth]{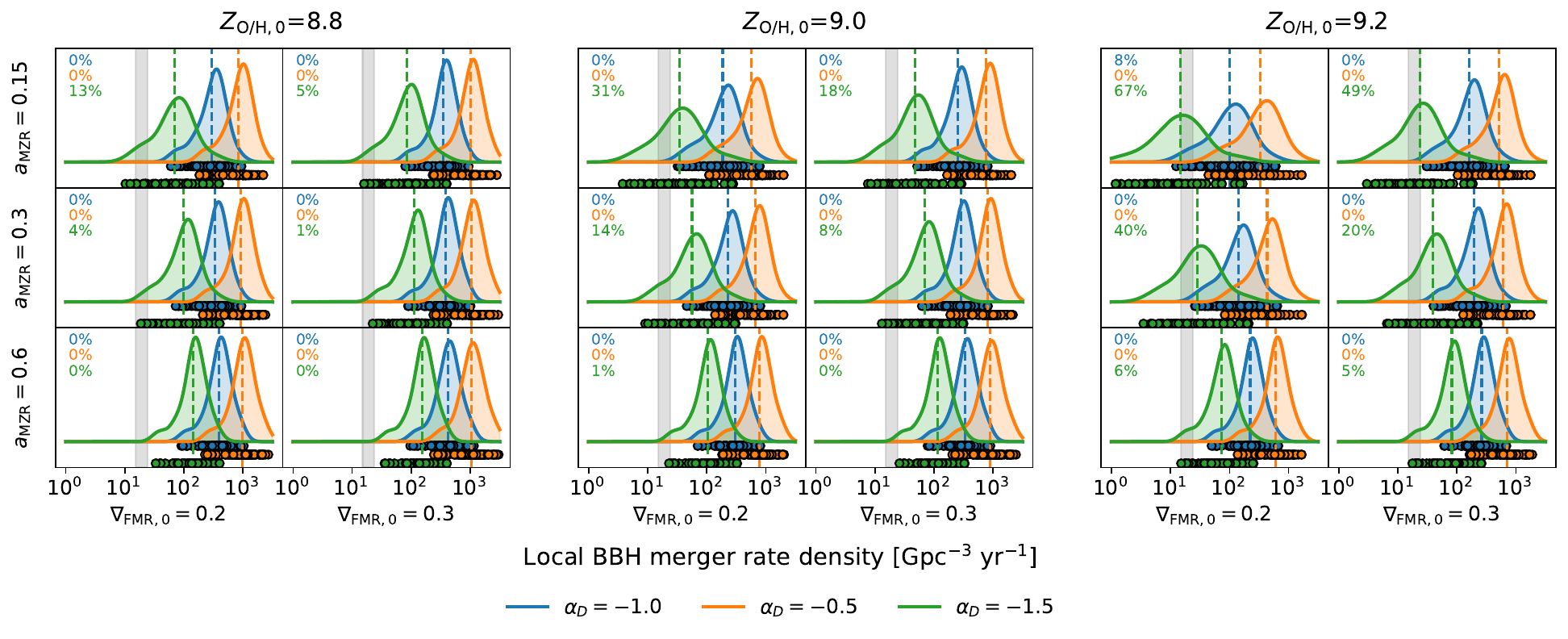}
    \caption{Same as Figure \ref{fig:main}, but here we show the effect of changing the exponent of the delay time distribution: $\alpha_D=-0.5$ (orange), $\alpha_D=-1.0$ (blue, fiducial), $\alpha_D=-1.5$ (green).}
    \label{fig:td}
\end{figure*}

Figure \ref{fig:td} shows the effect of the delay time distribution on our results. We vary the delay time exponent adopting different benchmark values: $\alpha_D=-0.5,\,-1,\,-1.5$, to flatten or steepen the distribution, and we recompute $\mathcal{R}_0$. Flattening the delay time distribution worsens the tension, as the local BBH merger rate receives more contribution from black holes formed at the cosmic noon, where there is the peak of the cosmic SFRD, or even higher redshift. Steepening the distribution, instead, helps in alleviating the tension, as less BBHs formed at high redshift contribute to the local BBH merger rate. In most of the metallicity variations, only a relatively small percentage of models ($<20\%$) fall within or below the LVK 90\% credible interval. However, in the most extreme metallicity variations, the percentage of models falling within or below the LVK 90\% credible interval rapidly increases, up to $67\%$ for the most favorable metallicity variation. This result confirms that binary evolution models favoring steeper delay time distributions may be one of the keys to reconcile the tension \citepalias{Boco2026}.

The strong effect of the delay time distribution on the predicted value of $\mathcal{R}_0$ motivates us to perform the self-consistent $t_d$ analysis for some of the models considered in this study.

\subsection{Self-consistent analysis}\label{sec:full}
The comprehensive analysis presented in the previous sections was meant to explore a wide variety of models, spanning different population synthesis codes, physical assumptions, parameter choices, and initial conditions. It was performed using the same delay time distribution, a simple power-law scaling $\propto t_d^{-1}$, for all the models. However, as shown in Section \ref{sec:delay}, the shape of the delay time distribution can have an important impact on the predicted BBH merger rate density. Thus, here we perform a self-consistent $t_d$ calculation of the local BBH merger density, using the delay time distributions directly obtained from the population synthesis simulations, for the subset of models yielding the lowest local BBH merger rates reported in their original work (see Section \ref{sec:methods}).

\begin{figure*}
    \centering
    \includegraphics[width=1.\linewidth]{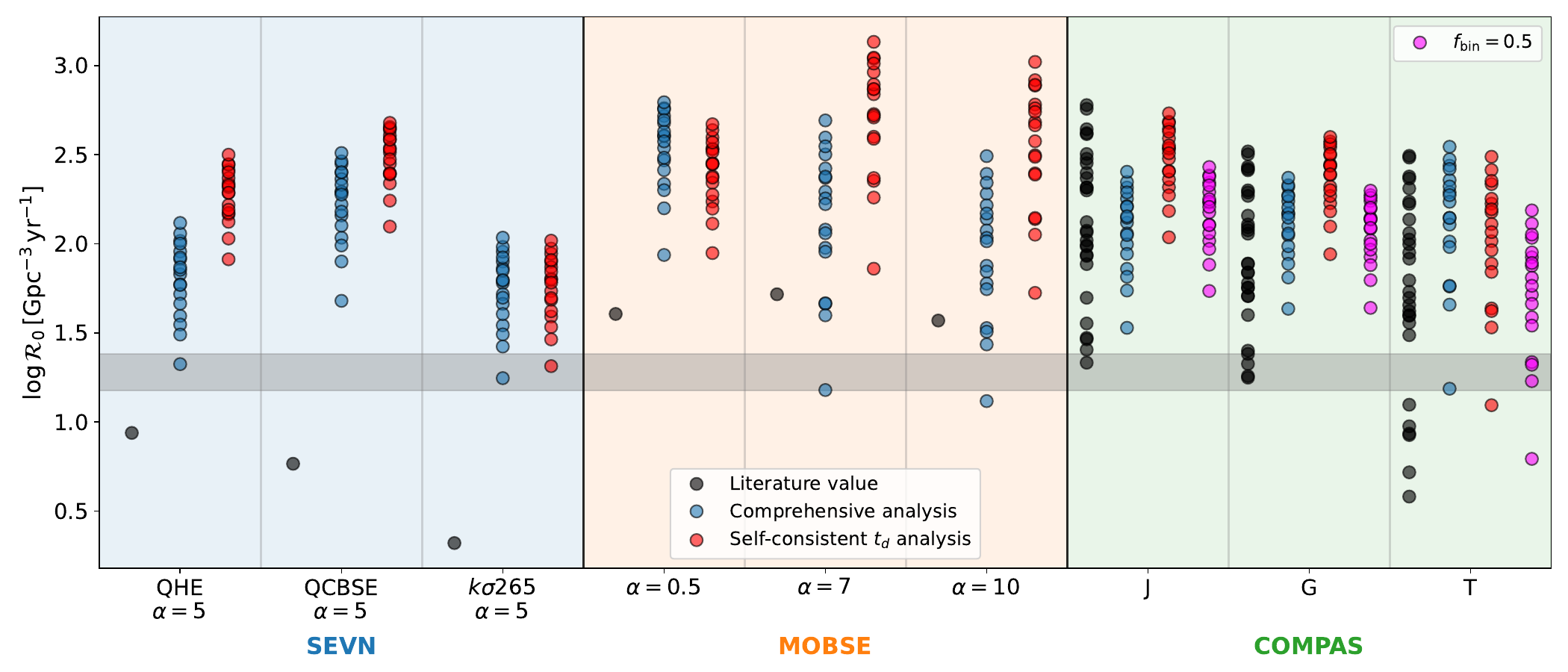}
    \caption{$\mathcal{R}_0$ values for the models predicting the lowest local BBH merger rate density. Each column represents one model. Comparison among literature values (black points), $\mathcal{R}_0$ derived from the comprehensive analysis (blue points), $\mathcal{R}_0$ derived from the self-consistent $t_d$ analysis (red points). Magenta points denote $\mathcal{R}_0$ obtained with \textsc{compas} and by assuming $f_\textrm{bin}=0.5$. Each of the points represents one of the 18 metallicity variations considered. The grey horizontal area is the 90\% credible interval for $\mathcal{R}_0$ estimated by LVK.}
    \label{fig:full-analysis}
\end{figure*}

The results are shown in Figure \ref{fig:full-analysis}. The self-consistent $t_d$ calculation (red points) leads to the same qualitative conclusion as the comprehensive analysis (blue points): adopting an observationally-derived SFRD shifts the predicted $\mathcal{R}_0$ toward higher values with repect to literature reported values (black points).

We note that the comprehensive analysis, based on a delay time distribution scaling as $t_d^{-1}$, yields not so different results from the self-consistent $t_d$ calculation. However, the tension with observations is generally even stronger when the self-consistent delay time distributions are adopted (red points versus blue points). This is because, for most models with low BBH merger efficiency, the delay time distribution tend to be flatter than $t_d^{-1}$ \citepalias[for more details, see the discussion in Section 4.2 of][]{Boco2026}. The \textsc{sevn} QHE model and the two \textsc{mobse} models with $\alpha_\textrm{CE}=7$ and $10$ can no longer match the observed $\mathcal{R}_0$ in any of the metallicity variations.

Only two models yield merger rates within or below the observational interval, and only for the most favorable metallicity variation. These are the \textsc{sevn} k$\sigma$265 model with strong natal kicks, and the \textsc{compas} model T, where the mass-loss rate of Wolf-Rayet winds has been multiplied by a factor of five.

Until now we reported our results considering the initial conditions and binary fractions for each model as they are in the original papers. In particular,  \citet{Broekgaarden2022} assume a binary fraction $f_\textrm{bin}=1$ in their simulation, while \citet{Iorio2023} and \citet{Santoliquido2021} adopt $f_\textrm{bin}=0.5$. Here, to facilitate a consistent comparison between models, we also show the \textsc{compas}  results renormalized by a factor $0.5$ (magenta points), which is a conservative lower limit. This shifts the $\mathcal{R}_0$ distribution downward by $0.3$ dex, making the \textsc{compas} model T within or below the LVK 90\% interval for four of the metallicity variations cosidered. The other \textsc{compas} models remain above the LVK inferred interval even after this renormalization.

\subsection{Caveats on the binary population-synthesis models}


In the comprehensive analysis we extended the investigation of \citet{Sgalletta2025} and \citetalias{Boco2026} by considering 101 binary evolution models from different codes (\textsc{sevn, mobse, compas}, and \textsc{stratrack}) with different physical assumptions and parameter's choices. We have shown that the low-metallicity tail of the SFRD$(z,Z)$ and the use of iron abundance shift most of the models to higher $\mathcal{R}_0$, generally above the LVK $90\%$ credible interval.

However, a few of the models analyzed here and/or reported by other studies \citep[see][for a review]{Broekgaarden2026b} yield particularly low $\mathcal{R}_0$ values compared to most others and might still remain within the LVK $90\%$ credible interval. Investigating these models in greater detail is particularly valuable, as they may help to identify the physical processes that can reduce the predicted merger rate sufficiently to reconcile population synthesis models with gravitational-wave data. For instance, such models involve strong stellar winds, high natal kicks, different formalisms for common-envelope evolution and mass-transfer stability. As shown in Section \ref{sec:full}, \textsc{compas} model T, which adopts Wolf-Rayet winds enhanced by a factor of 5, yields merger rates in agreement with observations for some metallicity variations. Similarly, \citet{Stevenson2022} presented some models with enhanced mass-loss rate that yield very low $\mathcal{R}_0$ values. Regarding natal kicks, very low BBH merger rates are obtained in models adopting the \citet{Hobbs2005} kicks without fallback \citep{Dominik2015, Mapelli2017,Belczynski2020,Sgalletta2025}, or with very high kick velocities, $\sigma_{\textrm{rms}}^\textrm{1D}=750$ km s$^{-1}$ \citep{Boesky2024b}. For mass transfer, models with only stable mass transfer \citep{vanson2022b} allow to substantially reduce merger rates \citep[see also][]{Sgalletta2025}. Finally, models with more restrictive mass-transfer instability criteria \citep{Olejak2021}, more conservative common envelope survival assumptions \citep{Romagnolo2025}, or a modified treatment of rejuvenation during mass transfer \citep{Ghodla2022} can predict rates in agreement with LVK, even when coupled to metallicity distributions featuring an extended low-metallicity tail.

\section{Conclusions}\label{sec:conclusions}
We have compared 101 BBH merger rate densities from isolated binary evolution models within a standardized cosmic star formation rate density and metallicity distribution. Our results can be summarized as follows.

\begin{itemize}
\item A metallicity-dependent cosmic star formation rate density derived from galaxy observations feature a significant low-metallicity tail, which can substantially increase the local BBH merger rates reported in literature.

\item Adopting an iron abundance-based metallicity distribution, which represents a stronger proxy than oxygen for the structure and evolution of $Z\gtrsim0.1~Z_\odot$ massive stars, leads to even more populated low-metallicity tails.

\item Using metallicity distributions consistent with the latest galaxy observations, leads only $\lesssim10\%$ of the binary evolution models considered to reproduce local BBH merger rates and only in the most favorable metallicity variation. The other metallicity variations result in BBH merger rate density above the $90\%$ credible interval inferred by LVK \citep{Abbott2026-gwtc5-population} for all the binary models considered.

\item In contrast, simplified symmetric metallicity distributions tend to underestimate the low-$Z$ tail if their $\log Z$ scatter is less than 0.6~dex, resulting in substantially lower predicted BBH merger rate densities.

\item The rates reported here account only for the isolated binary evolution channel. Additional channels (e.g., dynamical mergers in star clusters, AGN disks, or primordial black holes) should add to the rates predicted by isolated binary evolution, further increasing the tension between the data and the models.

\item The tension between models and gravitational-wave data can be alleviated only under possibly unrealistic assumptions on the starburst fraction ($f_\mathrm{SB}=0$), minimum galaxy mass ($10^8\, M_\odot$), and scatter around the fundamental metallicity relation, $\sigma_{\textrm{FMR}}=0.05$. Yet, even under these extreme assumptions, the tension with observations is not removed for most of the metallicity variations.

\item An alternative, more intriguing possibility is to consider binary evolution models predicting delay time distributions steeper than $t_d^{-1}$.

\item A further extension of the study to more population synthesis codes and different model variations is required. A first qualitative analysis shows that models with stronger Wolf-Rayet winds, higher natal kicks, or a different treatment of mass-transfer and common-envelope physics might be able to produce significantly lower $\mathcal{R}_0$ values \citep{vanson2025,Sgalletta2025,Broekgaarden2026b}.
\end{itemize}

Finally, comparing intrinsic properties of the BBH population, such as $\mathcal{R}_0$, without an analysis of all the other observables might lead to some misinterpretations and biases. For example, the merger rate overestimate may actually stem from an underestimate of the predicted mass function. A larger $\mathcal{R}_0$ with smaller average BH masses may lead to the same number of observed gravitational-wave events. Thus, in our future work we will include also other observed parameters (masses, spins, redshift evolution) in our analysis.

\begin{acknowledgements}
We thank the anonymous referee for critical reading and very helpful comments which improved a lot the quality of the manuscript. We thank Giuliano Iorio and Floor Broekgaarden for helpful discussions. LB acknowledges support by the Deutsche Forschungsgemeinschaft (DFG, German Research Foundation) in the form of a Walter Benjamin position -- Projektnummer 555003977. LB, MM, CS, and AR acknowledge financial support from the German Excellence Strategy via the Heidelberg Cluster of Excellence (EXC 2181 - 390900948) STRUCTURES. MM and AR acknowledge financial support from the European Research Council for the ERC Consolidator grant DEMOBLACK, under contract no. 770017 (PI: Mapelli). The authors also acknowledge support from the European Research Council for the ERC Consolidator grant IMBLACK, under contract no. 101197608 (PI: Mapelli). CS acknowledges financial support from the Alexander von Humboldt Foundation for the Humboldt Research Fellowship. MB acknowledges that this article was produced while attending the PhD program in PhD in Space Science and Technology at the University of Trento, Cycle XXXIX, with the support of a scholarship financed by the Ministerial Decree no. 118 of 2nd March 2023, based on the NRRP - funded by the European Union - NextGenerationEU - Mission 4 "Education and Research", Component 1 "Enhancement of the offer of educational services: from nurseries to universities” - Investment 4.1 “Extension of the number of research doctorates and innovative doctorates for public administration and cultural heritage” - CUP E66E23000110001 and support by the Italian grant Project SPACE-IT-UP by the Italian Space Agency and Ministry of University and Research, Contract Number 2024-5-E.0. We acknowledge support by the state of Baden-W\"urttemberg through bwHPC.
This research made use of \textsc{NumPy} \citep{Harris20}, \textsc{SciPy} \citep{SciPy2020}, \textsc{Matplotlib} \citep{Hunter2007}, and \textsc{sevn} (\href{https://gitlab.com/sevncodes/sevn}{https://gitlab.com/sevncodes/sevn}). We thank Lieke Van Son for the online material at \href{https://zenodo.org/records/14508864}{https://zenodo.org/records/14508864} that we used to compute $\mathcal{R}_0$ for different models, Giuliano Iorio for the online material at \href{https://zenodo.org/records/7794546}{https://zenodo.org/records/7794546} that we used to perform the self-consistent $t_d$ analysis for \textsc{sevn} models, Filippo Santoliquido, for sharing outputs of \textsc{mobse} models, and Floor Broekgaarden for the online material at \href{https://zenodo.org/records/5651073}{https://zenodo.org/records/5651073} that we used to perform the self-consistent $t_d$ analysis for \textsc{compas} models.
\end{acknowledgements}

\bibliographystyle{aa}
\bibliography{references}

\end{document}